\documentclass[preprint,12pt]{elsarticle}
\usepackage[T1]{fontenc}
\usepackage[utf8]{inputenc}
\usepackage{lmodern}
\usepackage{graphicx}
\usepackage{subfigure}
\usepackage{dcolumn}
\usepackage{bm}
\usepackage{amssymb}
\usepackage{amsfonts}
\usepackage{amsmath}
\usepackage{supertabular}
\usepackage{booktabs,longtable}

\journal{Physica A}

\begin{document}

\title{Complex network analysis of brain functional connectivity under a multi-step cognitive task}
\author[1,2]{Shi-Min Cai}
\ead{shimin.cai81@gmail.com}
\author[1,2]{Wei Chen}
\author[3]{Dong-Bai Liu}
\author[1,2]{Ming Tang}
\author[4]{Xun Chen}
\ead{xunchen@ece.ubc.ca}

\address[1]{Web Sciences Center, School of Computer Science and Engineering, University of Electronic Science and Technology of China, Chengdu 610073, P. R. China}
\address[2]{Big Data Research Center, University of Electronic Science and Technology of China, Chengdu 610073, P. R. China}
\address[3]{Department of Neurology, The Affiliated Jiangyin Hospital of Southeast University of Medical College, Jiangyin 214400, P. R. China}
\address[4]{Department of Electrical and Computer Engineering, University of British Columbia, Vancouver V6T 1Z4, BC, Canada}

\date{\today}

\begin{abstract}
Functional brain network has been widely studied to understand the relationship between brain organization
and behavior. In this paper, we aim to explore the functional connectivity of brain network under a
\emph{multi-step} cognitive task involving with consecutive behaviors, and further understand the effect of
behaviors on the brain organization. The functional brain networks are constructed base on a high
spatial and temporal resolution fMRI dataset and analyzed via complex network based approach.
We find that at voxel level the functional brain network shows robust small-worldness and scale-free
characteristics, while its assortativity and rich-club organization are slightly restricted to order
of behaviors performed. More interestingly, the functional connectivity of brain network in activated ROIs
strongly correlates with behaviors and behaves obvious differences restricted to order of behaviors performed.
These empirical results suggest that the brain organization has the generic properties of small-worldness
and scale-free characteristics, and its diverse function connectivity emerging from activated ROIs is
strongly driven by these behavioral activities via the plasticity of brain.
\end{abstract}


\maketitle

\section{Introduction}
Human brain, consisting of billions of neurons and synapses, is perhaps the most complex system ever known.
Its structural (or anatomic) and functional organization both behave complicated connectivity in the view
of graph and have been widely investigated via complex network theory in the neuroscience community. Plenty of works
focus on the topological properties of structural and functional brain networks derived from diffusion MRI,
funcitonal MRI (fMRI), electroencephalograph (EEG), magnetoencephalography (MEG), and multielectrode
array (MEA) data \cite{Bullmore2009complex,Sporns2012,Bullmore2012,Fallani2014,Zhang2014,Cui2016}. These
networks show both the generic small-worldness \cite{Watts1998} and scale-free characteristics \cite{Barabasi1999}
independent from the physiological and pathological states \cite{Eguiluz2005,He2007,Heuvel2008,Bassett2009,Stam2012}.
And, they also suggest high topological efficiency, robustness, modularity and rich club of hubs
\cite{Kasier2006,Achard2007,Meunier2009,Hagmann2008,Chavez2010,Heuvel2011,Zhuo2011,Sporns2016}.

It is also well known that human brain is physically expensive systems to built and run, that is, the adaptive
response (e.g., the capacity for information processing) of a brain network are constrained by its wiring costs.
~\cite{Kasier2006,Achard2007,Heuvel2009,Galos2012,Zalesky2014,Cole2014}. In other word,
these macro-scale functional response of human brain emerges from the synergistic dynamics
of micro-scale coupled neurons. Thus, the links between the brain structure and function can be
suggested by the neurons' dynamic activities. For examples, Honey \emph{et al} firstly use a
computational approach to relate the functional response of \emph{resting-state}
brain activity to the underlying structural connectivity and find there are structure-function correlations
at multiple temporal scales \cite{Honey2007}. Further, they demonstrate that although the resting-state functional
connectivity frequently exists between spatially distributed regions without direct structural edges,
its strength, persistence and spatial statistics are constrained by whole structural organization \cite{Honey2009}.
Similar result is also found in \cite{Cole2014,Cole2016}.

We have known that the structural and functional connectivity are characterized by common features,
and the functional connectivity is also correlated with the structural one even when the brain activity
evolves in the resting state. However, insofar as we know, how the multi-step behavioral activities
affect the emergence of functional connectivity and how its functional features involve with the structural
organization, have not yet been comprehensively studied by using complex network theory.
Herein, complex network based approach is applied to analyze the functional connectivity
when the brain activates under a multi-step cognitive task. The functional brain networks are
constructed at voxel and ROI (i.e., region of interests, functional area of structure organization)
 levels from the high spatially and temporally
resolved fMIR dataset. We find that at voxel level the functional brain network shows a number of the
statistic features, such as small-worldness, scale-free characteristics, assortativity
and rich-club organization, which are trivially affected by the order of behaviors performed.
More interestingly, at ROIs level, some statistic features of functional brain network are
obviously restricted to order of behaviors performed and correlated with these
activated ROIs. These empirical results suggest that the brain organization has some
generic properties and the diverse function connectivity emerging from activated ROIs
is strongly driven by these human behaviors via the plasticity of brain.

\section{Materials And Method}
\subsection{Materials}
The benchmark StarPlus fMRI dataset is collected by Just and his colleagues in
the $*$/$+$/$\$$ experiment at the Center for Cognitive Brain Imaging of Carnegie
Mellon University \cite{Carpenter1999,Wang2003,Mitchell2004}. In this cognitive
experiment, there are two different sessions for each individual subject.
The difference of the two sessions is the distinct order of behaviors performed
that involve with semantical and symbol stimulus. These two sessions are respectively
divided into four independent blocks, and each one was composed of a number of trials.
More specifically, each trial consisted of cognitive and rest segments.
In the cognitive segments of the first session, subjects are
presented with a sentence (semantic stimulus) on a screen such as ``It is
true that the star is bellow the plus'' for 4 seconds; then the sentence is replaced
with a black screen for another 4 seconds; finally subjects are shown with a picture
(symbol stimulus) depicting the geometric arrangement of the symbols $*$ and $+$,
and they should quickly judge whether the sentence describe the picture correctly
or not by pressing a button with the choice ``yes'' or ``no''. Once the judgment is
made or lasted more than 4 seconds, the picture would be removed from the screen.
Before repeating the next trial, there is a 15-second rest segment. The second session
has the similar procedure by simply switching the order of presenting sentence and the picture.

The fMRI images were collected every 0.5 second with the resolution $ 64 \times 64 \times 8 $.
Thus, there are around 54 images (~27 seconds) available for each trial. A total of twenty trials
are implemented for each subject in each session. We denote the two sessions as \emph{P}
(a picture presented before a sentence) and \emph{S} (a sentence shown before a picture), respectively.
For each session, the block was made up by several (4 or 8) continuous trials and thus have to contained rest
segments between adjacent trials.

Additionally, the cognitive experiments engaged several functional areas of
cerebral cortex (i.e., structural organization of brain), such as visual area for
the sentence/symbol reading (occipital lobe), spatial visualization
(inferior parietal sulcus, e.g. LIPS and RIPS) and recognizing (inferior temporal, e.g. LIT, RIT, LT and RT),
Broca area for language processing (left inferior frontal gyrus, e.g. LIFG),
Wernicke area for semantic analyzing (middle and superior temporal gyrus, angular gyrus, e.g. LTRIA and RTRIA),
motor area for the button pressing (supplementary motor area, e.g., SMA, LDLPFC and RDLPFC), etc.
Therefore, the voxels (i.e, 3-dimension pixels) of fMRI images were anatomically allocated into 25
ROIs. Based on that, the functional connectivity of brain network constructed
from these fMRI time series can be investigated at both small (voxel) and large (ROI) levels.

\subsection{Functional brain network construction}
The functional brain networks are extracted from the fMRI time series.
At time $t$, the fMRI images of brain activity are measured by a group of voxels. Thus, the voxel time
series $T(t,v)$ characterize the functional changes of cerebra cortex, and their correlations
imply the functional connectivity of areas of cerebra cortex. Taking these into consideration, it is natural to
define the nodes of functional brain network by these voxels, and associate their links with the correlations.
To encapsulate the temporal and spatial correlations, a synthetic measure is proposed here by coupling
the Pearson correlation and Euclidean distance. The calculation procedure will be described as follows.
For a block $x$ in session $P$, we first compute the Euclidean distances between each pair of the nodes, $v_{i}$ and $v_{j}$:
\begin{equation}
E_{x}(v_{i},v_{j})=\sqrt{\sum_{k=1}^{n}(T_{x}(t_{k},v_{i})-T_{x}(t_{k},v_{j}))^{2}},
\end{equation}
where $n$ denotes the length of voxel time series. Hence we get
\begin{equation}
\textit{E}_{p}(v_{i},v_{j})=\langle \textit{E}_{x}(v_{i},v_{j})\rangle  \hspace{1cm} \textit{x}\in P,
\end{equation}
where $\langle \cdot \rangle$ represents the mathematical expectation. $\textit{E}_{p}(v_{i},v_{j})$
determines the Euclidean distance between nodes $v_{i}$ and $v_{j}$. The larger Euclidean distance
suggests the weaker spatial correlation.

On the other hand, the Pearson correlation between each pair of nodes, $v_{i}$ and $v_{j}$ is computed as
\begin{equation}
\textit{R}_{x}(v_{i},v_{j})=\frac{\langle \textit{T}_{x}(t,v_{i})\textit{T}_{x}(t,v_{j})\rangle -\langle \textit{T}_{x}(t,v_{i})\rangle \langle \textit{T}_{x}(t,v_{j})\rangle }{\sigma (\textit{T}_{x}(v_{i}))\sigma (\textit{T}_{x}(v_{j}))},
\end{equation}
where $ \sigma ^{2}(\textit{T}_{x}(v))=\langle \textit{T}_{x}(t,v)^{2}\rangle-\langle \textit{T}_{x}(t,v)\rangle^{2} $.
And, we also get
\begin{equation}
\textit{R}_{p}(v_{i},v_{j})=\langle \textit{R}_{x}(v_{i},v_{j})\rangle \hspace{1cm} \textit{x}\in P.
\end{equation}

To characterize the Pearson correlation by a distance metric (termed as correlation distance),
an alternative approach satisfying the three axioms \cite{Cai2010} is
\begin{equation}
\textit{R}_{p}^{\prime }(v_{i},v_{j})=\sqrt{2(1-\textit{R}_{p}(v_{i},v_{j}))}.
\end{equation}
As $\textit{R}_{p}(v_{i},v_{j})$ ranges from [-1,1],  $\textit{R}_{p}^{\prime }(v_{i},v_{j})$ varies from [0,2].
It is emphasized that the smaller $\textit{R}_{p}(v_{i},v_{j})$, indicating weaker temporal correlation,
corresponds to the larger $\textit{R}_{p}^{\prime }(v_{i},v_{j})$. Thus, the correlation distance is
analogous to the Euclidean distance.

Finally, the synthetic measure of the correlation between each pair of nodes, $v_{i}$ and $v_{j}$,
is defined by the linear combination of correlation distance and Euclidean distance:
\begin{equation}
\textit{D}_{p}(v_{i},v_{j})=(1-\alpha)\textit{R}_{p}^{\prime }(v_{i},v_{j})+\alpha \textit{E}_{p}(v_{i},v_{j}).
\end{equation}
The coupling coefficient $\alpha $ makes $ \textit{E}_{p}(v_{i},v_{j}) $ and $ \textit{R}_{p}^{\prime }(v_{i},v_{j}) $
at the roughly same range and guarantees the balance between these two types of
distances since their probability distributions behave a similar shape \cite{Hong2012}.
For the session $S$, $D_{s}(v_{i},v_{j})$ can be obtained in the same way.



\begin{figure}
\center
\includegraphics[width=0.6\textwidth]{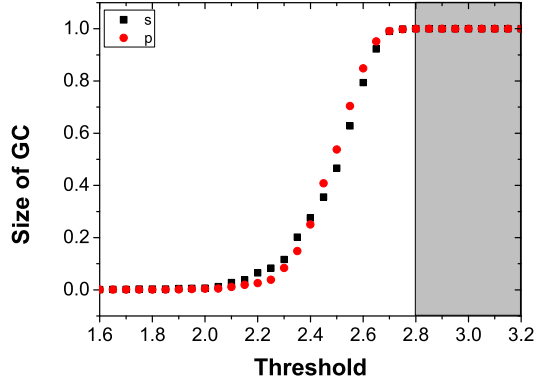}
\caption{\label{fig:threshold}(Color online) The relationship between the threshold and the size of giant component (GC). The size of GC is equal to 1 in gray area,
whose left edge of the gray area shows the point of percolation phase transition that is fixed as the threshold of binarization.}
\label{threshold}
\end{figure}

Although the synthetically measured matrix is able to characterize a weighted functional brain network, we binarize it into adjacent
one $A$ by choosing a proper threshold to simply and accurately quantify the functional connectivity of brain network.
The threshold is fixed by a percolation-based method that determines the size of giant component (GC), i.e., the largest connected subnetwork.
Figure \ref{threshold} shows that the size of GC changes as a function of threshold, suggesting that it gradually converges to 1 with increasing threshold.
And the point of percolation phase transition can be found at the left edge of the gray area in Fig. \ref{fig:threshold}, which
is properly fixed as the the threshold of binarization, $d_{c}=2.8$. Then, we deem that two nodes are functionally connected
(i.e., $a_{i,j}=1$, $a_{i,j} \in A$) if their distance $\textit{D}(v_{i},v_{j})$ does not exceed the threshold value $ d_{c} $.
Finally, a adjacent matrix is obtained to represent the functional brain network.

\subsection{Metrics}
Once the functional brain network is extracted, complex network theory,
as a common and effective method, is utilized to analyze the functional connectivity.
The measurements of topological structure are divided into voxel and ROI levels.
To keep our description as self-contained as possible,
a amount of important metrics should be reviewed briefly.

Degree $k_{i}$ of a node $i$ is the number of edges incident with the node, and is defined as
\begin{equation}
k_{i}=\sum_{j=1}^{N} a_{ij} \label{Degree},
\end{equation}
where $a_{ij}$ is an element of $A$ and $N$ is the total number of nodes. $\langle k \rangle$ thus denotes
the average degree (AD) of the whole network. The most basic topological feature of a network is the
degree distribution $P_{k}$, involved with the probability that a node is chosen randomly.

Clustering coefficient (CC) $ c_{i} $ of a node $i$ is defined as the ratio
of the number of edges between its neighbors ($E_{i}$) with respect to the total one as:
\begin{equation}
c_{i}=\frac{2E_{i}}{k_{i}(k_{i}-1)}.\label{CC}
\end{equation}
The CC of whole network, $C$, is the average value of all $ c_{i}$.

Path length (PL) $d_{ij}$ between two vertices $i$ and $j$ is the number of edges through the shortest path.
The average path length $L$ is calculated by averaging the path lengths of each pair of nodes as:
\begin{equation}
L=\frac{1}{\frac{1}{2}N(N+1)}\sum_{i\geq j}d_{ij}\label{PL}
\end{equation}

Assortativity coefficient (AC) $r$ is adopted to depict mixing pattern, and defined as,
\begin{equation}
r=\frac{\sum_{jk}jk(e_{jk}-q_{j}q_{k})}{\sigma_{q}^{2}}\label{Assortativity}.
\end{equation}
where $q_{k}=\frac{(k+1)P_{k+1}}{\sum_{j}jP_{j}}$ denotes the normalized distribution of the remaining degree,
$e_{jk}$ is defined as the joint probability distribution of the remaining degrees, and
$\sigma_{q}^{2}=\sum_{k}k^{2}q_{k}-[\sum_{k}kq_{k}]^{2}$ is the variance of the distribution $q_{k}$ \cite{Newman2002}.
For assortative mixing $r>0$, it suggests that nodes prefer to connecting with their similar one,
while for disassortative mixing $r<0$, it shows that
the connections more likely occur between dissimilar nodes \cite{Newman2003}.

The importance of a node in a network is usually characterized by a series of centrality indices. Besides node's degree, coreness and betweenness are often
employed. The coreness of a node $i$ is $k$ if the node $i$ belongs to $k$-core but doesn't exists $(k+1)$-core, where $k$-core is
the connected components of subgraph formed by repeatedly deleting all nodes with degree less than $k$ \cite{Gaertler2004}. Meanwhile,
betweenness of a node $i$ is defined as the sum of the shortest paths via node $i$~\cite{Mahadevan2005},
\begin{equation}
BC_i=\sum_{s,t \in N,s \neq i \neq t}d_{st}^{i}.\label{Betweenness}
\end{equation}

Average neighbor degree $k_{nn}^{i}$ of a node $i$ is used to describe degree-degree correlation
that associates with mixing pattern, and defined as:
\begin{equation}
k_{nn}^{i}=\sum_{j \in N}\frac{a_{i,j}k_j}{k_i}.\label{degree-degree corelation}
\end{equation}


The rich-club organization characterizes larger degree nodes' connectivity.
The rich-club coefficient $\varPhi(r/N)$ denotes the ratio of the number of existence edges $m$
with respect to the maximum number of possible edges $r(r-1)/2$
among those first $r$ nodes with the largest degree \cite{Zhou2004a,Zhou2004b}, which is described as
\begin{equation}
\varPhi(r/N)=\frac{2m}{r(r-1)}.\label{richclub}
\end{equation}

\section{Results and Discussion}

\subsection{Analysis of functional connectivity at voxel level}

Herein, the functional brain networks constructed from fMRI time
series at two sessions, $P$ and $S$, are concurrently analyzed to
investigate the effect of order of behaviors performed on functional connectivity.
Firstly, as shown in Tab. \ref{tab:table1}, the statistical features of
two functional brain networks are represented by generally characteristic parameters,
$C$, $\langle k \rangle$, $L$ and $r$. Both of them are
closely connected represented by large $\langle k \rangle$, and
the large $C$ and small $L$ show the robust small-worldness. However,
although $r > 0$ denotes assortative mixing of functional brain networks,
its strength differentiates from each other. These results straightly suggest
that the small-worldness independent from ordering of behaviors performed is
generic property of brain organization, yet to some extent the topological structures
of functional brain networks are also restricted to transient changes of behaviors.

\begin{longtable}{c c c c c c}
\hline
\hline
\textrm{}& \textrm{$N$}& \textrm{$\langle k \rangle$}& \textrm{$C$}& \textrm{$L$}& \textrm{$r$}\\
\hline
P & 4949 & 0.4220 & 129.2831 & 2.7340 & 0.1318 \\
\hline
S & 4949 & 0.4202 & 127.6193 & 2.7174 & 0.0545 \\
\hline
\hline
\caption{\label{tab:table1} The average of characteristic parameters for two functional brain networks.}
\end{longtable}



\begin{figure}
\center
\subfigure[]{\label{fig:degree_dsitribution}\includegraphics[width=0.5\textwidth]{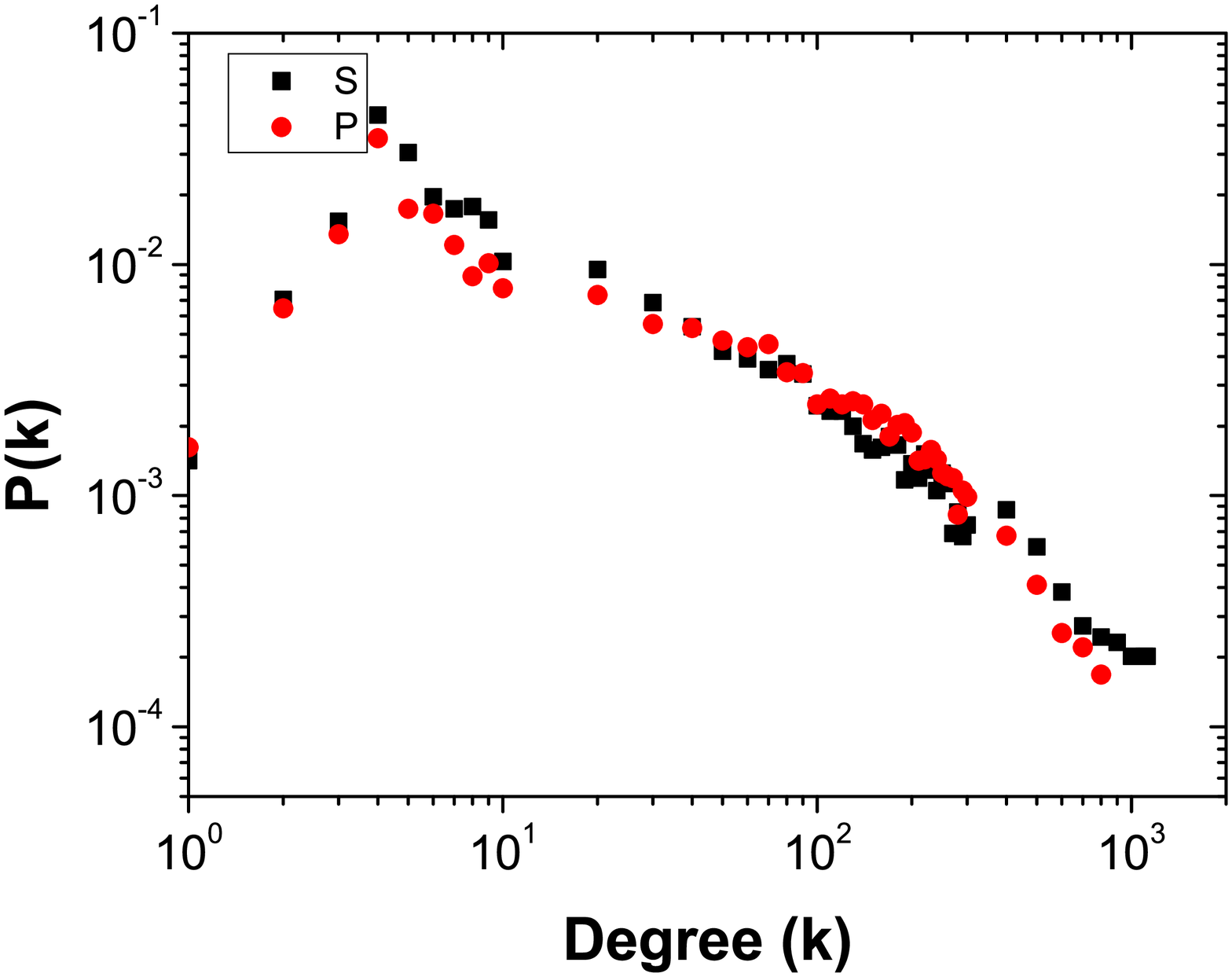}}
\subfigure[]{\label{fig:coreness}\includegraphics[width=0.45\textwidth]{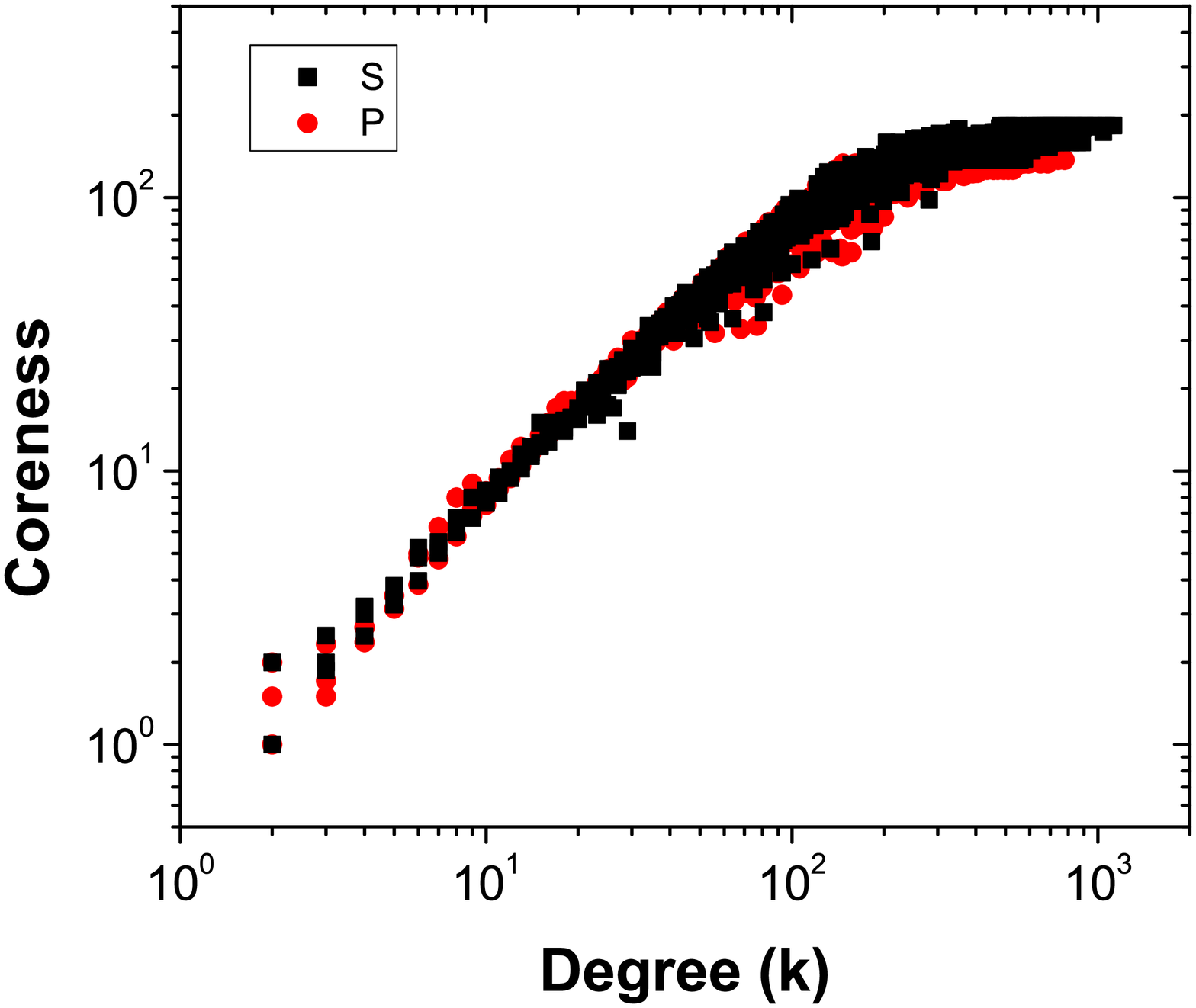}}
\subfigure[]{\label{fig:betweenness}\includegraphics[width=0.45\textwidth]{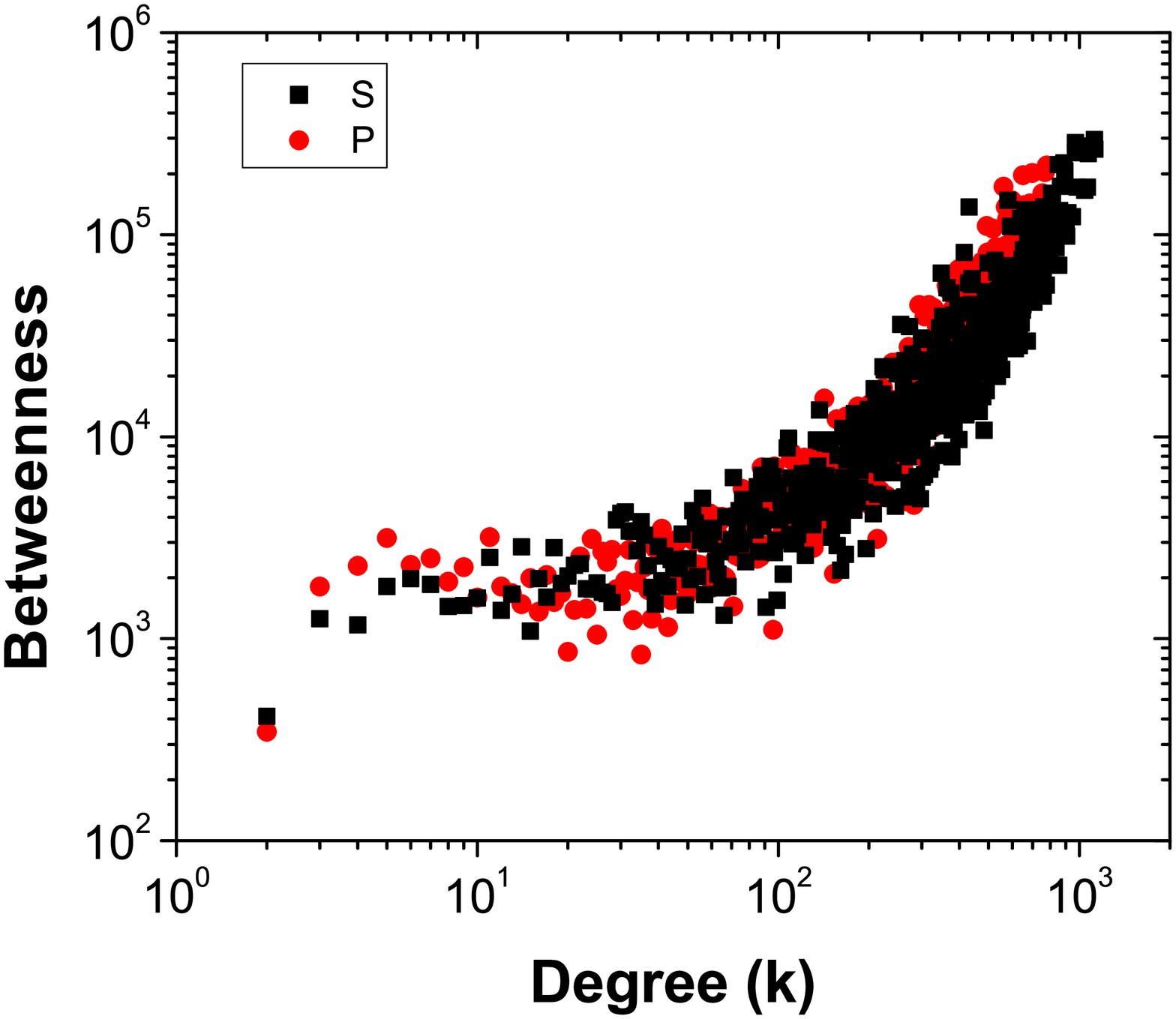}}
\subfigure[]{\label{fig:knn}\includegraphics[width=0.45\textwidth]{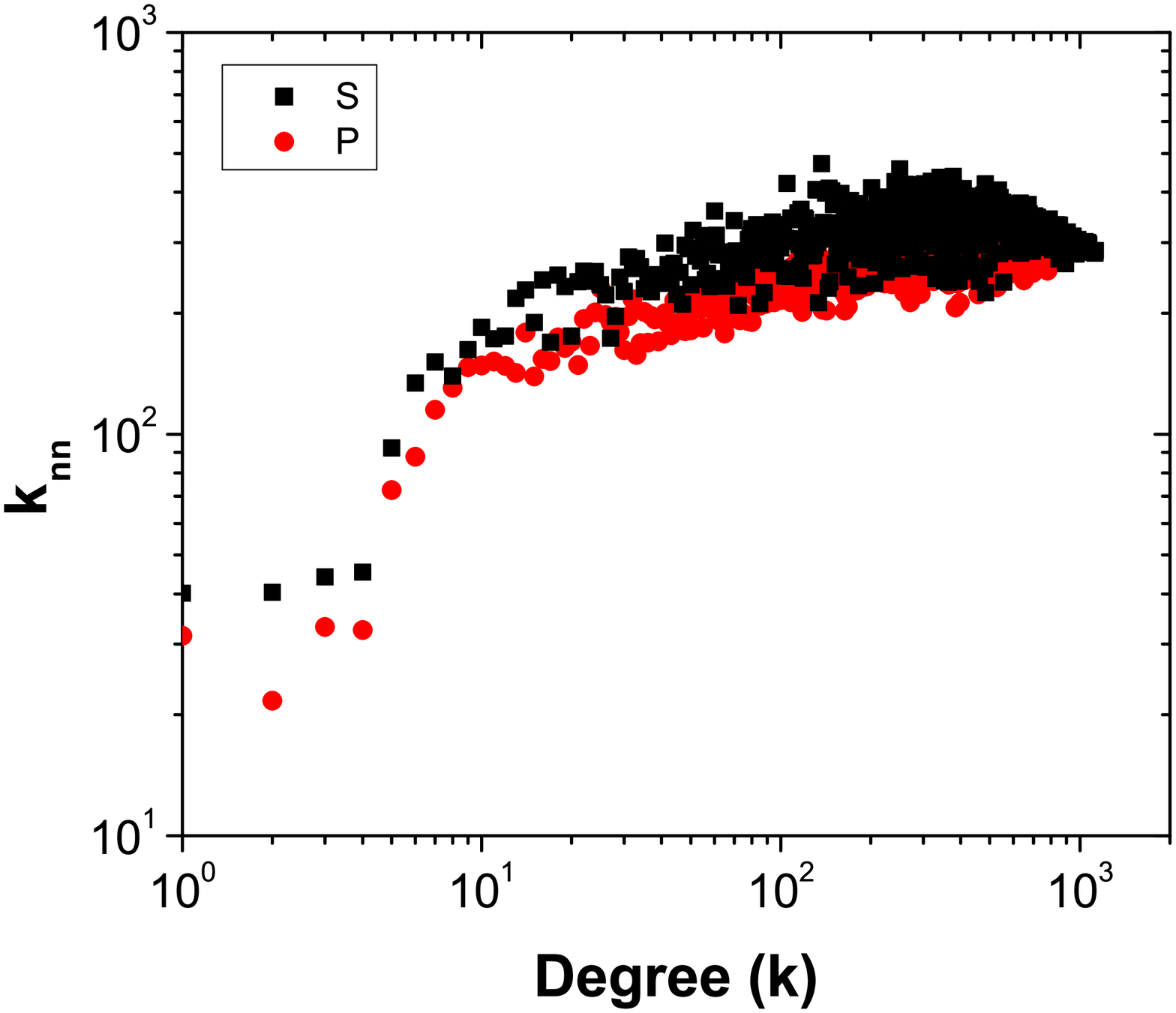}}
\caption{\label{fig:merge}(Color online) (a) Degree distributions at a log-log scale for two functional brain networks. Both of them have a power-law tail that suggests the scale-free characteristics. (b) Coreness as a function of degree $k$. The relation between coreness and degree for two functional brain networks both approximately obeys a power-law increasing trend when $k$ ranges from $10^{0}$ to $10^{2}$, but remains unchanged when $k$ exceeds $10^{2}$. (c) Betweenness as a function of degree $k$. The relation between betweeness and
degree for two functional brain networks both approximately keeps stable
when degree is less than $10^{2}$, but obeys a power-law increasing trend when $k$ exceeds $10^{2}$.
Note that both the coreness and betweenness centrality behave trivially different between \emph{P} and \emph{S}.
(d) Degree-degree correlation measured through $k_{nn}$ a function of $k$.
The positive correlation and the roughly different increasing slopes of relationship between
$k_{nn}$ and $k$ confirm the result obtained from assortativity coefficient.}
\end{figure}

Then, we analyze topological properties involving with nodes' degrees in details.
The degree distribution is a basic measurement. It is presented
at a log-log scale (using logarithmic bin) for two functional brain networks, respectively.
As shown in Fig.~\ref{fig:degree_dsitribution}, it can be seen that
the degree distributions are both approximately close to
a power law across multiple scales, which suggests the scale-free characteristics.
Moreover, the scale-free topological structure implies that there exists
hub-like nodes connecting with most of other ones and these nodes are
distributed into activated (or task-related) ROIs (see in Fig.~\ref{fig:localcoreness}).



The scale-free characteristics shows a heterogeneous nodes' connectivity of functional brain network.
Thus, these nodes behave different importance in the connectivity of functional brain network.
Here, we employ the coreness and betweenness to quantify node's importance, and mainly investigate
that they have a relation with degree. More concretely, the coreness indicates
the depth of node in functional brain network. Even if a node with a very high degree, its coreness may be very
small. Like a star network with $N$ nodes, the center node has a degree $N-1$ but its coreness is 0.
We compute each node's coreness according to \cite{Gaertler2004} and plot them as a function of degree $k$,
as shown in Fig.~\ref{fig:coreness}. It can be found that at these scales from $10^{0}$ to $10^{2}$ most of
nodes' coreness linearly increases with $k$, yet the nodes with degree larger than $10^2$ have
approximately similar coreness independent from $k$ and construct the nucleus of functional brain network.
Nevertheless, the relation between coreness and degree is trivially different between \emph{P} and \emph{S}.

And, the node's betweenness, describing its ability to control the information transitivity based on shortest path routing,
is also computed according to Eq.~\ref{Betweenness}. And, the betweeness as a function of $k$ is shown in
Fig.~\ref{fig:betweenness}. We can see that the values of betweeness are similar for these nodes with degree
less than $10^{2}$, then approximately increase according to a power law. Similarly, the relation between
betweenness and degree also behaves trivially different between \emph{P} and \emph{S}. However,
combining with the results in Fig.~\ref{fig:betweenness} and ~\ref{fig:coreness}, we can deduce that the nucleus of
functional brain network (e.g., these large degree nodes with $k>10^2$) controls and undertakes
the information transitivity of functional response and the functional brain networks are
easily affected by behavioral activities.

Furthermore, the degree-degree correlation measured through $k_{nn}$
as a function of $k$ reflect a potential mixing pattern.
The assortative mixing is denoted by the positive correlation that
$k_{nn}$ increases with $k$, while the disassortative mixing is
described by the negative correlation that $k_{nn}$ decreases with $k$.
Figure \ref{fig:knn} shows the relationship between $k_{nn}$ and $k$
for two functional brain networks. Obviously, both of them
are assortative mixing because $k_{nn}$ positively increase with
$k$, however, the increasing slopes are roughly different. These
results also correspond to assortativity coefficients in Tab. \ref{tab:table1}.
Note that the macroscopically quantitative difference of assortative mixing
for two functional brain networks may arise from the extent of activation
in visual and sematic areas. Thus, the order of behaviors performed doesn't
affect the assortative mixing of functional brain networks, but trivially alters
the assortativity coefficients.



\begin{figure}
\center
\includegraphics[width=0.6\textwidth]{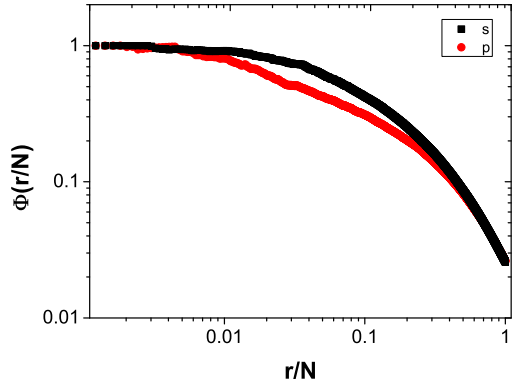}
\caption{\label{fig:richclub}(Color online) Rich-club coefficient at a double logarithmic scale for two functional brain networks.
It suggests that a few of larger degree nodes tend to densely connect with each other and form a stable rich club, yet
the nodes with median-scale degrees behaves different connectivity between \emph{P} and \emph{S}.}
\end{figure}

The rich-club organization indicates that there exist a few larger degree nodes in network,
and these nodes tend to densely connect with each other to constitutes ``rich club'' .
In \cite{Heuvel2011,Harriger2012,Collin2014}, they have demonstrated that
a small set of functional areas (i.e, nodes in functional brain network) are highly connected to form a dense rich club that paly a central role in global information integration. That is,
the existence of rich club organization in functional brain network is
due to the functional response of human brain. Herein, the rich-club coefficient is quantitatively shown in
Fig.~\ref{fig:richclub}. We can see that approximate 1$\%$ nodes are almost completely
connected regardless of the order of behaviors performed, however, these nodes with median-scale degrees
(e.g., $r/N$ $\in$ [0.01, 0.4]) behave different connectivity in restriction to the order of behaviors performed.
Thus, these results suggest that these nodes in rich club enable efficient information communication,
while those nodes with median-scale degrees play a non-trivial role in local information integration and correspond
to some functional areas related to behaviors. Furthermore, although the average degrees of two functional brain
networks are almost equal, the functional brain network \emph{P} has some larger degree nodes ($>10^3$) than
the functional brain network $S$ (see in Fig~\ref{fig:degree_dsitribution}). It leads to the difference of degree
heterogeneity between \emph{P} and \emph{S}. Thus, the difference of functional connectivity among the nodes
with median-scale degrees may arise from the degree heterogeneity.

\subsection{Analysis of functional connectivity at ROI level}

\begin{figure}
\center
\includegraphics[width=0.6\textwidth]{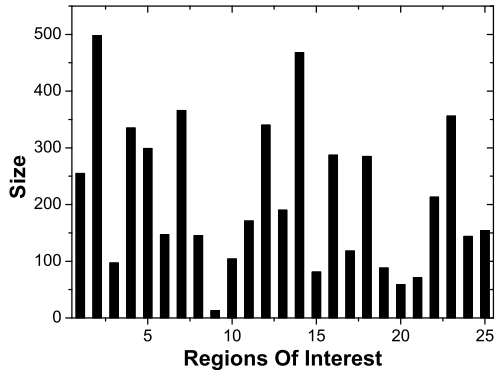}
\caption{\label{fig:size} Size of subnetworks at ROI level. These subnetworks show
the heterogeneous size and connectivity density. The horizontal axis from 1 to 25 represents the ROIs,
``CALC" ,``LDLPFC", ``LFEF", ``LIFG", ``LIPL", ``LIPS", ``LIT", ``LOPER", ``LPPREC", ``LSGA", ``LSPL", ``LT",
``LTRIA", ``RDLPFC", ``RFEF", ``RIPL", ``RIPS", ``RIT", ``ROPER", ``RPPREC", ``RSGA", ``RSPL", ``RT", ``RTRIA",
and ``SMA".}
\end{figure}

Since the voxels have been distributed into 25 anatomically defined ROIs as mentioned above,
it is interesting to probe the functional connectivity of brain network at ROI level. Specifically speaking,
we allocate the voxels (i.e., nodes) into 25 ROIs through ignoring external edges between ROIs, and obtain
25 isolated functional subnetworks. Note that there 22 ROIs are symmetrically distributed into
left and right hemisphere, respectively. Figure \ref{fig:size} shows that the sizes of subnetworks are heterogeneous,
of which the maximum is 498 and the minimum is 13, and approximately symmetrical for left and right hemisphere (except of ROI 9 and 20).
Note that the heterogenous subnetworks corresponding to ROIs arises from the difference of ROI's anatomic structure.
The horizontal axis labels 25 different ROIs, of which the order is in consistency with the following plots (see in Fig. \ref{fig:size}).

\begin{figure}
\center
\includegraphics[width=0.8\textwidth]{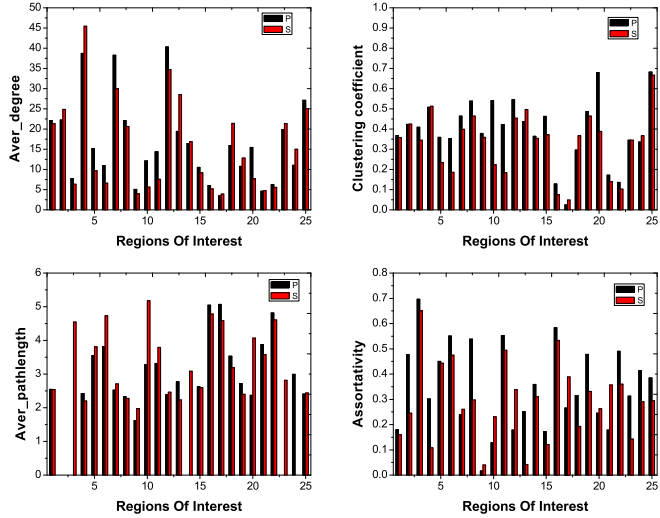}
\caption{\label{fig:local}(Color online)The statistical average values of four measurements, AD,
CC, PL and AC, in restricted to 25 ROIs (or subnetworks). Note that the
PL is equal to 0 due to incompletely connected subnetworks. The heterogeneous values of four
measurements suggest that the functional connectivity strongly correlates with ROIs, and
driven by behaviors.}
\end{figure}

For these 25 subnetworks, we also focus on the measurements
including AD, CC, PL, and assortativity. The statistical average values of four
measurements in restricted to $P$ and $S$ are shown in
Fig.~\ref{fig:local}, which suggests that the functional connectivity is different
between each pair of subnetworks, especially in these corresponding to activated ROIs.
More specifically, in the Fig.~\ref{fig:local}(a), the AD of subnetworks corresponding
to these activated ROIs 4, 7 (or 18), 12 (or 23), 25 is obviously larger than others,
and behave significant difference between $P$ and $S$ except of activated ROI 25. It
suggests that the connectivity of subnetworks strongly associates with
specific functional ares of cerebral cortex, and is obviously restricted to
order of behaviors performed. While in Fig.~\ref{fig:local}(b) and (c), they show the larger
CC and lower PL for all subnetworks regardless of behaviors, which suggests
the generic property of small-worldness. Due to the subnetworks aren't guaranteed to
completely connect, the PL misses in some of them. Moreover, in Fig.~\ref{fig:local} (d),
the assortativity suggests assortative mixing remaining in subnetworks although their values
are heterogeneous.

\begin{figure}
\center
\includegraphics[width=0.6\textwidth]{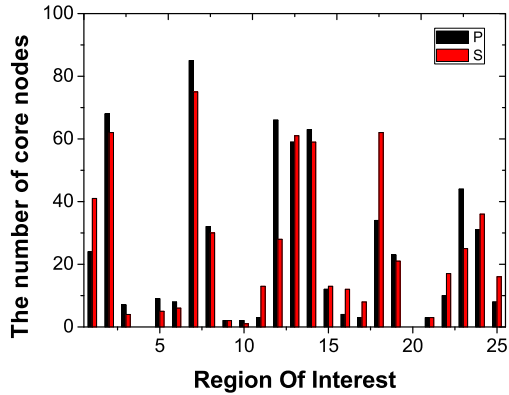}
\caption{\label{fig:localcoreness}(Color online) The most important 600 core nodes in term of coreness are distributed into
25 ROIs. We can find that these activated ROIs 2, 7, 12, 13, 14, 18, 23, 24 strongly associated with cognitive
function obviously include more core-like nodes, and some of them behave significant difference between \emph{P} and \emph{S}.}
\end{figure}

Nodes with high coreness are more important than those with low coreness as mentioned above.
We choose the most important 600 nodes in terms of the coreness and count how many vertices in each ROI,
to unveil which ROIs the nodes are mostly distributed in and whether the distribution has any differences in respective to
diverse cognitive states. Figure~\ref{fig:localcoreness} shows the number of core nodes in each ROI
restricted to \emph{P} and \emph{S}. It can be found that the ROI 7 (i.e., LIT) contains maximum nodes,
and most of them are distributed in temporal area (such like ROIs 12 (LT), 13 (LTRIA), 18 (RIT), 23 (RT), and 24 (RTRIA))
and central area (such like ROIs 2 ((LDLPFC) and 14 (RDLPFC)). Furthermore, the biggest difference for \emph{S} and \emph{P} is
also behaved in these ROIs 7, 12, 18, and 23. This suggests that the brain temporal lobes play a key role in cognition, and
when a subject judge whether a sentence described a picture correctly, the subject
may take more attention to the second stimulus, one explanation is that human
always has a deeper impression to the most recent stimulus.
As is well-known in literature, the temporal lobes are involved in the retention of visual
memories, processing sensory input, comprehending language, storing new memories, emotion, and senior
visual function (such as object recognition) \cite{Smith2006}. Thus, the empirical results are consistent
with previous findings.



\section{Conclusion}
The emphasis of this work unveils the association between brain organization and behavioral activities.
We systematically and comparatively analyze the function connectivity of brain network via complex network theory
when individual performs a consecutively cognitive task involving with diverse behaviors.
At voxel level, the functional brain network shows the generic properties of small-worldness
and scale-free characteristics, while its assortativity and rich club
organization are slightly restricted to the order of behaviors performed.
Furthermore, we divide the functional brain network into 25 subnetworks corresponding
to intra structural organization of ROIs, and find that these activated
ROIs associated with cognitive task obviously have larger AD and more core-like nodes.
Especially, these subnetworks corresponding to activated ROIs show that their
functional connectivity are restricted to the order of behaviors performed. Thus,
these empirical results suggest that the brain organization represented by the functional connectivity
are strongly driven by human behavioral activity (or cognitive task) via the plasticity of brain.


\section*{Acknowledgements}
This work is partially supported by the National Natural Science Foundation of China (Grant Nos. 61004002, 11575041, 61501164, 81571760 and 61673086)
and the Fundamental Research Funds of the Central Universities (Grant No. ZYGX2015J153).



\begin{thebibliography}{99}

\bibitem{Bullmore2009complex} E. T. Bullmore and O. Sporns, Nat. Rev. Neurosci. 10, 186 (2009)

\bibitem{Sporns2012} O. Sporns, NeuroImage 62, 881 (2012)

\bibitem{Bullmore2012} E. T. Bullmore and O. Sporns, Nat. Rev. Neurosci. 13, 336 (2012)

\bibitem{Fallani2014} F. De Vico Fallani, J. Richiardi, M. Chaavez, and S. Archard, Phil. Trans. R. Soc. B 369, 20130521 (2014)


\bibitem{Cui2016} H. Cui, et al. Hum. Brain Mapp. 37, 1459 (2016)

\bibitem{Zhang2014} J. Zhang, K. M. Kendrick, G. Lu, and J. Feng, Cereb. Coretex, doi:10.1093/cercor/bhu173


\bibitem{Watts1998} D. J. Watts and S. H. Strogatz, Nature 393, 440 (1998)

\bibitem{Barabasi1999} A. L. Barab\'{a}si and R. Albert, Science 286, 509 (1999)



\bibitem{Eguiluz2005} V. M. Eguiluz, D. R. Chialvo, G. A. Cecchi, M. Baliki, and A. V. Apkarian, Phys. Rev. Lett. 94(1), 018102 (2005).

\bibitem{He2007} Y. He, Z. J. Chen, A. C. Evans, Cereb. Coretex 17, 2407 (2007).

\bibitem{Heuvel2008} M. P. van den Heuvel, C. J. Stam, M. Boersma, and H. E. Hulshoff Pol, NeuroImage 43, 528 (2008).

\bibitem{Bassett2009} D. S. Bassett and E. T. Bullmore, Curr. Opin. Neurol. 22, 340 (2009)

\bibitem{Stam2012} C. J. Stam and E. C. W. van Straaten, Clin. Neruophysiol. 123, 1067 (2012)









\bibitem{Kasier2006} M. Kasier and C. C. Hilgetag, PLoS Comput. Biol. 2, e95 (2006)

\bibitem{Achard2007} R. Achard and E. T. Bullmore, PLoS Comput. Biol. 3, e1 (2007)

\bibitem{Hagmann2008} P. Hagmann, L. Cammoun, X. Gigandet, R. Meuli, C. J. Honey, V. J. Wedeen and O, Sporns, PloS Biol. 6, e159 (2008)

\bibitem{Meunier2009} D. Meunier, R. Lambiotte, A. Fornito, K. D. Ersche and E. T. Bullmore, Front. Neuroinform 3, 37 (2009)


\bibitem{Chavez2010} M. Chavez, M. Valencia, V. Navarro, V. Latora and J. Martinerie, Phys. Rev. Lett. 104, 118701 (2010)

\bibitem{Heuvel2011} M. P. van den Heuvel and O. Sporns, J. Neurosci. 31, 15775 (2011).

\bibitem{Zhuo2011} Z. Zhuo, S. M. Cai, Z. Q. Fu, J. Zhang, Phys. Rev. E 84, 031923 (2011).

\bibitem{Sporns2016} O. Sporns and R. F. Betzel, Ann. Rev. Psychol. 67(19), 19 (2016).




\bibitem{Heuvel2009} M. P. van den Heuvel, C. J. Stam, R. Kahn and H. E. Hulshoff Pol, J. Neurosci. 29, 7619 (2009).

\bibitem{Galos2012} L. K. Galos, H. A. Makse and M. Sigman,  Proc. Natl. Acad. Sci. USA 109, 2825 (2012).

\bibitem{Zalesky2014} A. Zalesky, A. Fornito, L. Cocchi, L. L. Gollo, and M. Breakspear, Proc. Natl. Acad. Sci. 111(28), 10341 (2014).

\bibitem{Cole2014} M. W. Cole, D. S. Bassett, J. D. Power, T. S. Braver, S. E. Petersen, Neuron 83, 238 (2014).



\bibitem{Honey2007} C. J. Honey, R. K\"{o}tter, M. Breakspear, and O Sporns, Proc. Natl. Acad. Sci. 104(24), 10240 (2007).

\bibitem{Honey2009} C. J. Honey, O. Sprons, L. Cammoun, X. Gigandet, J. P. Thiran, R. Meuli, and P. Hagmann, Proc. Natl. Acad. Sci. 106(6), 2035 (2009).


\bibitem{Cole2016} M. W. Cole, T. Ito, T. S. Braver, Cereb. Coretex, doi: 10.1093/cercor/bhv072



\bibitem{Carpenter1999} P. A. Carpenter, M. A. Just, T. A. Keller, W. F. Eddy, and K. R. Thuborn, NeuroImage 10, 216 (1999).

\bibitem{Wang2003} X. Wang, R. Hutchinson, and T. Mitchell, Training fMRI Classifiers to Detect Cognitive States across Multiple Human Subjects, in Proceedings of NIPS'03.

\bibitem{Mitchell2004} T. M. Mitchell, R. Hutchinson, R. S. Niculescu, F. pereira, X. Wang, M. Just, and S. Newman, Mach. Lern. 57, 145 (2004).


\bibitem{Hong2012} L. Hong, S. M. Cai, J. Zhang. Z. Zhuo, Z. Q. Fu, and P. L. Zhou, Chaos 22, 033128 (2012).


\bibitem{Cai2010} S. M. Cai, Y. B. Zhou, T. Zhou, and P. L. Zhou, Int. J. Mod. Phys. C 21(03), 433 (2010).

\bibitem{Newman2002} M. E. J. Newman, Phys. Rev. Lett. 89(20), 208701 (2002).

\bibitem{Newman2003} M. E. J. Newman, Phys. Rev. E 67(2), 026126 (2003).

\bibitem{Gaertler2004} M. Gaertler and M. Patrignani, Dynamic analysis of the autonomous system graph, Proceedings of International Workshop on Inter-domain Performance and Simulation (IPS 2004), Budapest, Hungary, pp.13-24 (2004).

\bibitem{Mahadevan2005} P. Mahadevan. D. Krioukov, M. Fomenkov, B. Huffaker, X. Dimitropoulos, K. Claffy, and A. Vahdat, arXiv:cs/0508033.

\bibitem{Zhou2004a} S. Zhou and R. J. Mondrag\'{o}n, IEEE Commun. Lett. 8(3), 180 (2004).

\bibitem{Zhou2004b} S. Zhou and R. J. Mondrag\'{o}n,  Phys. Rev. E 70(6), 066108 (2004).





\bibitem{Harriger2012} L. Harriger, M. P. van den Heuvel, and O. Sporns, PLoS ONE 7(9), 446497 (2012).

\bibitem{Collin2014} G. Collin, O. Sporns, R. C. W. Mandl, and M. P. van den Heuvel, Cereb. Cortex 2(9), 2258 (2014).

\bibitem{Smith2006} E. E. Smith and S. M. Kosslyn, Cognitive Psychology: Mind and Brain, New Jersey: Prentice Hall (2006).




\end{thebibliography}
\end{document}